# Quantum generalized Reed-Solomon codes: Unified framework for quantum MDS codes

Zhuo Li, Li-Juan Xing, and Xin-Mei Wang

State Key Laboratory of Integrated Service Networks, Xidian University, Xi'an,

Shannxi 710071, China

We construct a new family of quantum MDS codes from classical generalized Reed-Solomon codes and derive the necessary and sufficient condition under which these quantum codes exist. We also give code bounds and show how to construct them analytically. We find that existing quantum MDS codes can be unified under these codes in the sense that when a quantum MDS code exists, then a quantum code of this type with the same parameters also exists. Thus as far as is known at present, they are the most important family of quantum MDS codes.

## I. INTRODUCTION

The quest to build a scalable quantum computer that is resilient against decoherence errors and operational noise has sparked a lot of interest in quantum error-correcting codes [1-11]. The early research has been confined to the study of binary quantum error-correcting codes, but more recently the theory was extended to nonbinary codes that are useful in the realization of fault-tolerant computations.

Among these codes quantum MDS codes are optimal in the sense that the minimum distance is maximal, since they meet the quantum Singleton bound [10]. Recently many families of quantum MDS codes have been found by various different approaches [12-15].

In this paper we derive a new family of quantum MDS codes that are based on classical generalized Reed-Solomon codes [16]. We call them quantum generalized Reed-Solomon (QGRS) codes. Firstly we give the definition of QGRS code. Then we give the necessary and sufficient condition for the existence of QGRS code, which shows that the problem of finding QGRS codes can be transformed into the problem of finding the weight distributions of certain classical codes. So it is possible to search for codes of this type. But this is not practical for large codes. Thus we also show how to construct QGRS codes analytically in the end.

Another achievement of this paper is the unification of various quantum MDS codes under QGRS codes. Recall that various existing quantum MDS codes have been found by different approaches, which brings inconvenience for application. Occurrence of QGRS codes changes this situation. Each preceding quantum MDS code has a counterpart in QGRS codes. In the other word, it will turn out that in all the known cases, when a quantum MDS code exists, then a QGRS code with the same parameters also exists. Thus as far as is known at present, QGRS codes are the most important family of quantum MDS codes.

## II. PRELIMINARIES

Firstly, let us recall the definition and property of classical generalized RS codes briefly [16].

Let $\mathbf{a} = (\alpha_1, \ldots, \alpha_n)$ where the $\alpha_i$ are distinct elements of $GF(q^m)$, and let $\mathbf{v} = (v_1, \ldots, v_n)$ where the $v_i$ are nonzero elements of $GF(q^m)$. Then the generalized RS code, denoted by $C_{GRS}(\mathbf{a}, \mathbf{v}, k)$, consists of all vectors $(v_1 F(\alpha_1), v_2 F(\alpha_2), \ldots, v_n F(\alpha_n))$ where $F(z)$ is any polynomial of degree $< k$ with coefficients from $GF(q^m)$.

Let $C_{GRS}(\mathbf{a}, \mathbf{v}, k)$ be an $[n, k, n-k+1]_q$ generalized RS code. The $[n+1, k, n-k+2]_q$ extended generalized RS code, $C^*_{GRS}(\mathbf{a}, \mathbf{v}, k)$, has generator matrix

$$\mathbf{G}^* = \begin{bmatrix} v_1 & \cdots & v_n & 0 \\ \alpha_1 v_1 & \cdots & \alpha_n v_n & 0 \\ \alpha_1^2 v_1 & \cdots & \alpha_n^2 v_n & 0 \\ \cdots & \cdots & \cdots & \cdots \\ \alpha_1^{k-1} v_1 & \cdots & \alpha_n^{k-1} v_n & 1 \end{bmatrix}.$$

When $k = 3$ and $q$ even, one more parity check can always be added, producing an $[n+2, 3, n]_q$ extended generalized RS code, $C^*_{GRS}(\mathbf{a}, \mathbf{v}, 3)$, by using the generator matrix

$$\mathbf{G}^* = \begin{bmatrix} v_1 & \cdots & v_n & 0 & 0 \\ \alpha_1 v_1 & \cdots & \alpha_n v_n & 1 & 0 \\ \alpha_1^2 v_1 & \cdots & \alpha_n^2 v_n & 0 & 1 \end{bmatrix}.$$

Additionally, a typical method which we will use for construction is as follows.

***Lemma 1 [11].*** Let $C$ be an $[n, k]_{q^2}$ classical code contained in its Hermitian dual, $C^{\perp_h}$, such that $d = \min\{wt(C^{\perp_h} \setminus C)\}$. Then there exists an $[\![n, n-2k, d]\!]_q$ quantum code.

# III. QUANTUM GENERALIZED RS CODES

Now we start to show our contributions.

*Definition 2.* Quantum generalized Reed-Solomon (QGRS) code is the quantum code that is derived from classical Hermitian self-orthogonal generalized RS or extended generalized RS code by lemma 1.

Let $C$ be a $GF(q^2)$-linear code. Define $P(C) = \langle (c_i d_i^q)_{i=1}^n : \mathbf{c}, \mathbf{d} \in C \rangle^{\perp} \cap GF(q)^n$, which is equivalent to the puncture code introduced by Rains [10].

*Theorem 3.* A QGRS code $[\![r, r-2k, k+1]\!]_q$ exists if and only if there exists a codeword of weight $r$ in $P(C_{GRS}^*(\boldsymbol{\alpha}, \mathbf{1}, k))$ where vector $\boldsymbol{\alpha} = (\alpha_1, \ldots, \alpha_{q^2})$ contains all elements of $GF(q^2)$ and $\mathbf{1}$ denotes vector of 1's.

*Proof.* (Only if.) Suppose a QGRS code $[\![r, r-2k, k+1]\!]_q$ exists, and $C$ is the corresponding $[r, k, r-k+1]_{q^2}$ Hermitian self-orthogonal generalized RS code with generator matrix

$$\mathbf{G} = \begin{bmatrix} v_1 & \cdots & v_r \\ \alpha_{i_1} v_1 & \cdots & \alpha_{i_r} v_r \\ \cdots & \cdots & \cdots \\ \alpha_{i_1}^{k-1} v_1 & \cdots & \alpha_{i_r}^{k-1} v_r \end{bmatrix}.$$

Then $\sum_{j=1}^{r} (\alpha_{i_j}^s v_j)(\alpha_{i_j}^t v_j)^q = \sum_{j=1}^{r} \alpha_{i_j}^{s+tq} v_j^{q+1} = 0$ for $0 \leq s, t \leq k-1$. Setting the coordinate $i_j$ equal to $v_j^{q+1}$ for $1 \leq j \leq r$ and the others to $0$ gives a codeword of weight $r$ in $P(C_{GRS}^*(\boldsymbol{\alpha}, \mathbf{1}, k))$. The case of extended generalized RS code is handled in the same way. The proof of the converse is similar. Q.E.D.

This theorem tells us that the problem of finding QGRS codes can be transformed into the problem of finding the weight distributions of certain classical codes. Once we find out the weight distribution of $P(C^*_{GRS}(\boldsymbol{\alpha},\mathbf{1},k))$, we can derive all possible QGRS codes by theorem 3.

*Corollary 4.* (Bounds.) Let $\mathcal{L}$ be an $[\![n, n-2k, k+1]\!]_q$ QGRS code. Then:

(i) If $q \geq 3$, $k \leq q$;

(ii) For all $k$ and $q$ except for $k=3$ and $q$ even, $n \leq q^2 + 1$;

(iii) For $k=3$ and $q$ even, $n \leq q^2 + 2$.

*Proof.* (i) Suppose $k \geq q+1$, then $P(C^*_{GRS}(\boldsymbol{\alpha},\mathbf{1},k))$ has parity check matrix

$$\begin{bmatrix} 1 & \cdots & 1 & 0 \\ \alpha_1 & \cdots & \alpha_{q^2} & 0 \\ \cdots & \cdots & \cdots & \cdots \\ \alpha_1^{q^2-1} & \cdots & \alpha_{q^2}^{q^2-1} & 0 \\ \alpha_1^{(k-1)(q+1)} & \cdots & \alpha_{q^2}^{(k-1)(q+1)} & 1 \end{bmatrix}$$

whose determinant is nonzero. Thus there exists no QGRS code by theorem 3. (ii) and (iii) are obvious by theorem 3. Q.E.D.

In the next section, we will show that QGRS code $[\![q^2+1, q^2-2q+1, q+1]\!]_q$ exists. For $k=3$ and $q=2,4$, we also find QGRS codes $[\![6,0,4]\!]_2$ and $[\![18,12,4]\!]_4$. Hence the bounds given by corollary 4 are tight.

## IV. ANALYTICAL CONSTRUCTION FOR QGRS CODES

Theorem 3 makes it possible to search for QGRS codes. But this is not practical for large codes. In this section we show how to construct QGRS codes analytically. We begin with a lemma.

**Lemma 5.** Let $\alpha_1, \ldots, \alpha_n$ be $n$ distinct elements of any field. Then we have

$$\sum_{i=1}^{n} \frac{\alpha_i^h}{\prod_{\substack{j=1 \\ j \neq i}}^{n}(\alpha_i - \alpha_j)} = 0$$

for $h \leq n-2$.

*Proof.* It is easy to verify that

$$\sum_{i=1}^{n} \frac{\alpha_i^h}{\prod_{\substack{j=1 \\ j \neq i}}^{n}(\alpha_i - \alpha_j)} = \begin{vmatrix} 1 & 1 & \cdots & 1 \\ \alpha_1 & \alpha_2 & \cdots & \alpha_n \\ \cdots & \cdots & \cdots & \cdots \\ \alpha_1^{n-2} & \alpha_2^{n-2} & \cdots & \alpha_n^{n-2} \\ \alpha_1^h & \alpha_2^h & \cdots & \alpha_n^h \end{vmatrix} \Bigg/ \begin{vmatrix} 1 & 1 & \cdots & 1 \\ \alpha_1 & \alpha_2 & \cdots & \alpha_n \\ \cdots & \cdots & \cdots & \cdots \\ \alpha_1^{n-2} & \alpha_2^{n-2} & \cdots & \alpha_n^{n-2} \\ \alpha_1^{n-1} & \alpha_2^{n-1} & \cdots & \alpha_n^{n-1} \end{vmatrix}.$$ Q.E.D.

Now we can give the main theorem of this paper:

**Theorem 6.** There exist QGRS codes $[\![n, n-2k, k+1]\!]_q$ where code parameters satisfy $\begin{cases} n = q^2 + 1 \\ k = q \end{cases}$, $\begin{cases} n = q^2 - l \\ k \leq q - l - 1 \\ 0 \leq l \leq q-2 \end{cases}$, $\begin{cases} n = mq - l \\ k \leq m - l \\ 0 \leq l < m \\ 1 < m < q \end{cases}$, or $\begin{cases} n \leq q \\ k \leq \lfloor n/2 \rfloor \end{cases}$.

*Proof.* (Constructive proof.) Let $GF(q) = \{\beta_1, \ldots, \beta_{q-1}, \beta_q = 0\}$, and let $\{1, \gamma\}$ be the basis of $GF(q^2)/GF(q)$. Then

$$GF(q^2) \triangleq \{\alpha_{iq+j} \mid \alpha_{iq+j} = \beta_{i+1}\gamma + \beta_j, 0 \leq i \leq q-1, 1 \leq j \leq q\},$$

$$\prod_{i=1}^{q}(\alpha_{sq+j} - \alpha_{tq+i}) = \prod_{i=1}^{q}[(\beta_{s+1}\gamma + \beta_j) - (\beta_{t+1}\gamma + \beta_i)] = (\beta_{s+1} - \beta_{t+1})\prod_{i=1}^{q}(\gamma + \beta_i), \quad s \neq t, \quad (1)$$

$$\prod_{\substack{i=1 \\ i \neq j}}^{q}(\alpha_{sq+j} - \alpha_{sq+i}) = \prod_{\substack{i=1 \\ i \neq j}}^{q}[(\beta_{s+1}\gamma + \beta_j) - (\beta_{s+1}\gamma + \beta_i)] = \prod_{i=1}^{q-1}\beta_i. \quad (2)$$

When $n = q^2 + 1$, set $\boldsymbol{\alpha} = (\alpha_1, \alpha_2, \ldots, \alpha_{q^2})$. It is easy to verify that $C^*_{GRS}(\boldsymbol{\alpha}, \mathbf{1}, q)$ is Hermitian self-orthogonal. Thus QGRS code $[\![q^2+1, q^2-2q+1, q+1]\!]_q$ exists.

When $n = q^2 - l$ with $0 \leq l \leq q-2$, set $\boldsymbol{\alpha} = (\alpha_1, \alpha_2, \ldots, \alpha_{q^2-l})$ and set $\mathbf{v} = (v_1, v_2, \ldots, v_{q^2-l})$ where $v_i = \prod_{j=q^2-l+1}^{q^2} (\alpha_i^q - \alpha_j^q)$. Then for $C_{GRS}(\boldsymbol{\alpha}^q, \mathbf{v}, k)$ and $C_{GRS}(\boldsymbol{\alpha}, \mathbf{1}, n-k)$,

$$\sum_{i=1}^{n}(v_i \alpha_i^{qs})^q \alpha_i^t = \sum_{i=1}^{n} \prod_{j=q^2-l+1}^{q^2}(\alpha_i - \alpha_j)\alpha_i^{s+t} = (\prod_{j=1}^{q^2-1}\alpha_j)\sum_{i=1}^{n} \frac{\prod_{j=q^2-l+1}^{q^2}(\alpha_i - \alpha_j)\alpha_i^{s+t}}{\prod_{j=1}^{q^2-1}\alpha_j}$$

$$= (\prod_{j=1}^{q^2-1}\alpha_j)\sum_{i=1}^{n} \frac{\prod_{j=q^2-l+1}^{q^2}(\alpha_i - \alpha_j)\alpha_i^{s+t}}{\prod_{\substack{j=1 \\ j \neq i}}^{q^2}(\alpha_i - \alpha_j)} = (\prod_{j=1}^{q^2-1}\alpha_j)\sum_{i=1}^{n} \frac{\alpha_i^{s+t}}{\prod_{\substack{j=1 \\ j \neq i}}^{n}(\alpha_i - \alpha_j)} = 0$$

for $s \leq k-1$, $t \leq n-k-1$ by lemma 5. So $C_{GRS}(\boldsymbol{\alpha}^q, \mathbf{v}, k)$ is Hermitian dual to $C_{GRS}(\boldsymbol{\alpha}, \mathbf{1}, n-k)$. Now if $k \leq q-l-1$, $C_{GRS}(\boldsymbol{\alpha}^q, \mathbf{v}, k) \subseteq C_{GRS}(\boldsymbol{\alpha}, \mathbf{1}, n-k)$.

When $n = mq - l$ with $1 < m < q$, $0 \leq l < m$, from (1) and (2), there is a nonzero constant $\lambda_i \in GF(q)$, with $1 \leq i \leq mq$, such that $\prod_{\substack{j=1 \\ j \neq i}}^{mq}(\alpha_i - \alpha_j) = \lambda_i \zeta^{m-1}$ where $\zeta = \prod_{i=1}^{q}(\gamma + \beta_i)$. Set $\boldsymbol{\alpha} = (\alpha_1, \alpha_2, \ldots, \alpha_{mq-l})$, $\boldsymbol{\mu} = (\mu_1, \ldots, \mu_{mq-l})$, and $\mathbf{v} = (v_1, \ldots, v_{mq-l})$ where $\mu_i = v_i \prod_{j=mq-l+1}^{mq}(\alpha_i^q - \alpha_j^q)$, $v_i^{q+1} = \lambda_i^{-1}$. Then for $C_{GRS}(\boldsymbol{\alpha}^q, \boldsymbol{\mu}, k)$ and $C_{GRS}(\boldsymbol{\alpha}, \mathbf{v}, n-k)$,

$$\sum_{i=1}^{n}(\mu_i \alpha_i^{qs})^q (v_i \alpha_i^t) = \sum_{i=1}^{n} v_i^{q+1} \alpha_i^{s+t} \prod_{j=mq-l+1}^{mq}(\alpha_i - \alpha_j) = \zeta^{m-1} \sum_{i=1}^{n} \frac{\alpha_i^{s+t} \prod_{j=mq-l+1}^{mq}(\alpha_i - \alpha_j)}{\lambda_i \zeta^{m-1}}$$

$$= \zeta^{m-1} \sum_{i=1}^{n} \frac{\alpha_i^{s+t} \prod_{j=mq-l+1}^{mq}(\alpha_i - \alpha_j)}{\prod_{\substack{j=1 \\ j \neq i}}^{mq}(\alpha_i - \alpha_j)} = \zeta^{m-1} \sum_{i=1}^{n} \frac{\alpha_i^{s+t}}{\prod_{\substack{j=1 \\ j \neq i}}^{n}(\alpha_i - \alpha_j)} = 0$$

for $s \leq k-1$, $t \leq n-k-1$ by lemma 5. So $C_{GRS}(\boldsymbol{\alpha}^q, \boldsymbol{\mu}, k)$ is Hermitian dual to $C_{GRS}(\boldsymbol{\alpha}, \mathbf{v}, n-k)$. Now if $k \leq m-l$, $C_{GRS}(\boldsymbol{\alpha}^q, \boldsymbol{\mu}, k) \subseteq C_{GRS}(\boldsymbol{\alpha}, \mathbf{v}, n-k)$.

When $n \leq q$, set $\boldsymbol{\alpha} = (\beta_1, \beta_2, \ldots, \beta_n)$ and $\mathbf{v} = (v_1, \ldots, v_n)$ where $v_i^{q+1} = 1 / \prod_{\substack{j=1 \\ j \neq i}}^{n} (\beta_i - \beta_j)$. Then for $C_{GRS}(\boldsymbol{\alpha}, \mathbf{v}, k)$ and $C_{GRS}(\boldsymbol{\alpha}, \mathbf{v}, n-k)$,

$$\sum_{i=1}^{n} (v_i \beta_i^s)^q (v_i \beta_i^t) = \sum_{i=1}^{n} v_i^{q+1} \beta_i^{s+t} = \sum_{i=1}^{n} \frac{\beta_i^{s+t}}{\prod_{\substack{j=1 \\ j \neq i}}^{n}(\beta_i - \beta_j)} = 0$$

for $s \leq k-1$, $t \leq n-k-1$ by lemma 5. So $C_{GRS}(\boldsymbol{\alpha}, \mathbf{v}, k)$ is Hermitian dual to $C_{GRS}(\boldsymbol{\alpha}, \mathbf{v}, n-k)$. Now if $k \leq \lfloor n/2 \rfloor$, $C_{GRS}(\boldsymbol{\alpha}, \mathbf{v}, k) \subseteq C_{GRS}(\boldsymbol{\alpha}, \mathbf{v}, n-k)$.  Q.E.D.

In theorem 6 we just give the smallest range of code parameters among which QGRS codes exist. We have proved that the range can be enlarged slightly. This is to say, more QGRS codes can be constructed analytically.

## V. DISCUSSION

Now let us study existing quantum MDS codes. For quantum MDS codes $[\![5,1,3]\!]_q$ in [12], there exist QGRS codes with parameters $[\![5,1,3]\!]_q$ by theorem 6. For quantum MDS codes $[\![6,2,3]\!]_p$ and $[\![7,3,3]\!]_p$ ($p \geq 3$) in [13], by direct search and theorem 6 we can find QGRS codes with corresponding parameters easily. For quantum MDS codes $[\![n, n-2d+2, d]\!]_q$ for $n \leq q$, and $[\![q^2-s, q^2-2d+2-s, d]\!]_q$ for some s in [14], corresponding QGRS codes can be obtained from theorem 6 and theorem 3 respectively. For quantum MDS codes $[\![q^2-q\alpha, q^2-q\alpha-2v-2, v+2]\!]_q$ for $0 \leq v \leq q-2$ and $0 \leq \alpha \leq q-v-1$ in [15], QGRS codes with corresponding parameters can be derived from theorem 6 immediately.

To sum up, in all the known cases, various quantum MDS codes can be unified under QGRS codes. When a quantum MDS code exists, then a QGRS code with the same parameters also exists. Thus as far as is known at present, QGRS codes are the most important family of quantum MDS codes.